\documentclass{article}[12pt]
\usepackage{amsfonts}

\newbox\strutbox
\setbox\strutbox=\hbox{\char"B}
\def\strut{\relax\ifmmode\copy\strutbox\else\unhcopy\strutbox\fi}
\def\ialign{\everycr{}\tabskip0pt\halign}
\def\eqalign#1{\null \,\vcenter {\openup\jot \mathsurround 0pt
    \ialign{\strut \hfil$\displaystyle{##}$&$\displaystyle
      {{}##}$\hfil\crcr#1\crcr}}\,}

\def\eqalignno#1{\tabskip 0pt plus 1 fill \halign to\displaywidth{\hfil$\tabskip0pt\everycr{}\displaystyle{##}$\tabskip 0pt &$\tabskip0pt\everycr{}\displaystyle {{}##}$\hfil \tabskip 0pt plus 1 fill&\llap {$\tabskip0pt\everycr{} ##$}\tabskip 0pt \crcr #1\crcr }}

\begin{document}
\title{Quantum Amplitudes in Black-Hole Evaporation
  I. Complex Approach}
\author{A.N.St.J.Farley and P.D.D'Eath\protect\footnote{Department of Applied Mathematics and Theoretical Physics,
Centre for Mathematical Sciences,
University of Cambridge, Wilberforce Road, Cambridge CB3 0WA,
United Kingdom}}

\maketitle

\begin{abstract}
This paper is the first in a project concerned with the
quantum-mechanical decay of a Schwarzschild-like black hole,
formed by gravitational collapse, into almost-flat space-time and 
weak radiation at a very late time.  The approach taken here is, 
in fact, applicable to a much wider class of quantum calculations 
than those concerning gravitational collapse, including quantum
properties of a variety of cosmological examples involving weak 
anisotropic perturbations.  In this work, we are concerned with 
evaluating quantum amplitudes (not just probabilities) for transitions 
from initial to final states.  The present quantum description shows 
that no information is lost in collapse to a black hole.  In a general 
asymptotically-flat context (not necessarily involving local 
gravitational collapse to a black hole), one may specify a quantum 
amplitude by posing boundary data on (say) an initial space-like 
hypersurface $\Sigma_I$ and a final space-like hypersurface 
$\Sigma_F{\,}$, together with the Lorentzian proper-time interval $T$
which separates them, as measured near spatial infinity.  Suppose,
for simplicity, that the Lagrangian contains Einstein gravity with 
only a minimally-coupled massless scalar field $\phi{\,}$, describing
the matter present.  Then the boundary data can (for example) be chosen 
to be $h_{ij}$ and $\phi$ on the two hypersurfaces, where 
$h_{ij}{\,}(i,j=1,2,3)$ is the intrinsic spatial 3-metric.  
The classical boundary-value problem, corresponding to the calculation 
of this quantum amplitude, is badly posed, being a boundary-value
problem for a wave-like (hyperbolic) set of equations.  Following 
Feynman's $+i\epsilon$ prescription, one makes the problem well-posed 
by rotating the asymptotic time-interval $T$ into the complex: 
$T\rightarrow{\mid}T{\mid}\exp(-i\theta)$, 
with ${\,}0<\theta\leq\pi/2{\,}$.  After calculating the amplitude for 
$\theta >0{\,}$, one then takes the 'Lorentzian limit'
$\theta\rightarrow 0_{+}{\,}$.  All calculations in this work are 
based on this procedure.  For example, in the following Paper II,
this will be used to calculate amplitudes for given final weak
configurations of the scalar field, representing scalar radiation 
on the final hypersurface $\Sigma_F$ at a late time $T{\,}$.
\end{abstract}

\begin{section}{Introduction}
This paper is the first in a project concerned with the calculation
of quantum amplitudes (not just probabilities) associated with quantum
fields (including gravity), in the case that strong gravitational
fields may be present; this may be compared with the effective-action 
approach of [1,2]. 
The most obvious example -- the original
motivation for this work -- concerns quantum radiation associated
with gravitational collapse to a black hole [3-11].  But the
framework adopted here is more general, and certainly does not depend 
on whether there is a classical Lorentzian-signature collapse to a 
black hole.  It includes the case of local collapse which is not
sufficient to lead to (Lorentzian) curvature singularities, and also
quantum processes in cosmology, where, for example, anisotropies in
the Cosmic Microwave Background Radiation can be computed, and depend
crucially on the underlying Lagrangian for gravity and matter [12].

To exemplify the underlying ideas, let us consider the case of local 
collapse (whether or not to a black hole).  Thus, the gravitational 
field is taken to be asymptotically flat.  Suppose, as in this paper, 
that we consider Einstein gravity coupled minimally to a massless
scalar field $\phi{\,}$.  In classical gravitation, we are used to 
describing this by means of a Cauchy problem, giving evolution to the 
future (say) of an initial spacelike hypersurface ${\cal S}$, which 
extends to spatial infinity.  Denoting by $h_{ij}$ the intrinsic 
spatial metric on ${\cal S}\; (i,j=1,2,3)$, Cauchy data would, 
loosely speaking, consist of $h_{ij}{\,},\phi$ and the corresponding 
normal derivatives on ${\cal S}$.  By contrast, in quantum theory, 
one typically asks for the amplitude to go from an initial
configuration such as $(h_{ij},\phi)_I$ on an initial hypersurface 
$\Sigma_I{\,}$, to a final configuration $(h_{ij},\phi)_F$ on a 
final hypersurface $\Sigma_F{\,}$.  The problem of finding the 
quantum amplitude should (naively) be completely posed, once one 
has also specified the (Lorentzian) proper-time interval, measured 
orthogonally between the surfaces $\Sigma_I$ and $\Sigma_F$ 
near spatial infinity.

Much of the 'non-intuitive' nature of quantum mechanics can be traced
to the 'boundary-value' nature of such a quantum amplitude [13], as
compared with the familiar classical initial-value problem.  A crucial
mathematical aspect of this difference, responsible for a good part of
the 'non-intuition', is that the {\it classical} version of the
problem of calculating a quantum amplitude, as posed above, would involve
solving the classical field equations (typically hyperbolic), subject
to the given boundary data $(h_{ij},\phi)_{I,F}$ on surfaces separated
near spatial infinity by a Lorentzian time interval $T{\,}$.  As is well
known, a boundary-value problem for a hyperbolic equation is typically
not well posed.  For typical boundary data, a classical solution will
not exist [14,15]; or, if it does exist, it will be non-unique.
The straightforward cure for this ill, due to Feynman [13], is of
course to rotate the Lorentzian time-interval $T$ into the complex:
$T\rightarrow{\mid}T{\mid}\exp(-i\theta){\,}$, 
with ${\,}0<\theta\leq\pi/2{\,}$.

Consider, for a simple example, a Dirichlet boundary-value problem,
but where one begins with the 'Riemannian' version, for the Laplace 
equation (rather than the wave equation) in Euclidean 2-space 
${\Bbb E}^{2}{\,}$, with data given on an initial boundary 
$\{\tau =0\}$ and on a final boundary $\{\tau =T\}{\,}$.  
Thus, one considers a scalar field $\phi(\tau{\,},x)$, obeying
$${{{\partial}^{2}\phi}\over{{\partial}{\tau}^{2}}}{\;}
+{\;}{{{\partial}^{2}{\phi}}\over{{\partial}x^{2}}}{\;}{\,}
={\;}{\,}0{\quad};{\qquad}{\quad}0<{\tau}<T{\;},{\quad}
-\infty <x< +\infty{\;},\eqno(1.1)$$
on the assumption that $\phi$ decays rapidly as 
${\mid}x{\mid}{\,}{\rightarrow}{\,}{\infty}{\,}$.  Suppose, 
for simplicity, that the Dirichlet boundary data are taken to be 
of the form
$$\phi(\tau =0{\,},x){\;}{\,}={\;}{\,}0{\;},
{\qquad}{\quad}\phi(\tau =T{\,},x){\;}{\,}={\;}{\,}{\phi}_{1}(x){\;}.
\eqno(1.2)$$
This problem is treated simply, by taking a Fourier transform with
respect to $x{\,}$.  Define, for example,  
$${\Phi}_{1}(k){\;}{\,}={\;}{\,}{{1}\over{\sqrt{2\pi}}}{\,}
{\int}_{-\infty}^{\infty}{\,}e^{-ikx}{\;}{\phi}_{1}(x){\;}dx{\;}.
\eqno(1.3)$$ 
Then the (unique) solution is given by
$$\phi(\tau{\,},x){\;}{\,}
={\;}{\,}{{1}\over{\sqrt{2\pi}}}{\,}
{\int}_{-\infty}^{\infty}{\,}e^{ikx}{\;}
{{\sinh(k\tau)}\over{\sinh(kT)}}{\;}{\Phi}_{1}(k){\;}dk{\;}.\eqno(1.4)$$
As with a solution of any elliptic partial-differential equation with
analytic coefficients, $\phi$ is automatically a (real- or complex-) 
analytic function of both arguments $\tau$ and $x{\,}$ [14].

One may then, as above, rotate the 'time-interval' $T$ into the
complex: ${\,}T\rightarrow T\exp(i\theta){\,}$, where 
${\,}0\leq{\,}\theta <\pi/2{\,}$.  The integral expression (1.4) 
continues to give the (unique) solution to the differential equation, 
where it is understood that the boundary data (1.2) are left invariant 
as $\theta$ is varied.  This follows since (provided 
${\,}0\leq{\,}\theta <\pi/2{\,}$) the denominator in the integrand of
Eq.(1.4):
$$\sinh\bigl(kTe^{-i\theta}\bigr){\;}{\,}
={\;}{\,}\cos\bigl(kT\sin\theta \bigr){\,}\sinh\bigl(kT\cos\theta\bigr)
{\,}-{\,}i{\,}\sin\bigl(kT\sin\theta\bigr){\,}
\cosh\bigl(kT\cos\theta\bigr){\;},\eqno(1.5)$$ 
is non-zero for all real $k\neq 0{\,}$, whence the integrand is smooth
for all real $k{\,}$.

This good behaviour of our linear boundary-value example, in the complex
case ${\,}0<{\,}\theta<\pi/2{\,}$, also follows from more general
arguments [16].  For given $\theta{\,}\in{\,}(0,\pi/2){\,}$, let us
define a new 'time' coordinate ${\,}y=\tau\exp(i\theta){\,}$, which is
adapted to the new 'time-interval' $T\exp(i\theta){\,}$.  In terms of
the new coordinates $(y{\,},x){\,}$, the Laplace equation (1.1) reads
$$e^{2i\theta}{\,}{{{\partial}^{2}\phi}\over{\partial y^{2}}}{\;}
+{\;}{{{\partial}^{2}\phi}\over{\partial x^{2}}}{\;}{\,}
={\;}{\,}0{\quad},\eqno(1.6)$$
and the boundary conditions read
$$\phi(y=0{\,},x){\;}{\,}={\;}{\,}0{\;},
{\qquad}{\quad}\phi\bigl(y=Te^{i\theta},x\bigr){\;}{\,}
={\;}{\,}\phi_{1}(x){\quad}.\eqno(1.7)$$
The potential $\phi(y{\,},x)$ is a complex solution of Eq.(1.6), which
is a {\it strongly elliptic} partial differential equation in the
sense of [16].  The property of strong ellipticity guarantees
existence and uniqueness in this linear example.

But, in the Lorentzian case $\theta =\pi/2{\,}$, the denominator
becomes $-{\,}i\sin(kT){\,}$, which has zeros at $k=n\pi/T{\,}$, 
for $n$ any integer, positive or negative.  Typical boundary data
$\phi_{1}(x)$ will not have ${\Phi}_{1}(n\pi/T)=0$ for a single value
of $n$ ($n$ integer, $n\neq 0$).  Following this argument, one can
show that there is, in general, no solution to the Lorentzian 
(wave-equation) boundary-value problem above.  Thus, in this simple
example, one can already see that the Lorentzian Dirichlet
boundary-value problem is badly posed.

In our coupled non-linear gravitational/scalar-field example, the 
extreme case $\theta =\pi/2$ would correspond to a purely Euclidean 
time-interval ${\mid}T{\mid}{\,}$, and classically one would then 
be solving the field equations for Riemannian gravity with a scalar 
field $\phi{\,}$.  Since these field equations are 
'elliptic {\it modulo} gauge' -- see [16] -- one would expect to have
a well-posed 
boundary-value problem, with existence and uniqueness.  The
intermediate case ${\,}0<\theta<\pi/2{\,}$ requires the interval
$T$ and any classical solution to involve the complex numbers
non-trivially.  If the problem turns out to be strongly elliptic, up
to gauge,
then the complex case ${\,}0<\theta<\pi/2{\,}$ would again be expected 
to have the good existence and uniqueness properties of the real 
elliptic case.

In practice, one typically treats the case in which the gravitational 
and scalar initial and final data are close to spherical symmetry.  
Hence, as a leading approximation, one begins by studying the 
spherically-symmetric Einstein-scalar system.  This is treated in [17] 
for Lorentzian signature and is outlined in [18] for Riemannian signature.  
In the Riemannian case, the metric is taken (without loss) in the form 
(3.5,6) below, involving two functions $a,b,$ which depend on two
coordinates $\tau{\,},r{\,}$.  For Riemannian signature, the field 
equations are given in Eqs.(3.7-11) below, as partial differential 
equations for the metric functions $a,b$ and the scalar field 
$\phi{\,}$, with respect to $\tau$ and $r{\,}$.  Even in the 
spherically-symmetric case, very little is known rigorously about 
existence and uniqueness for the Riemannian (or complex) 
boundary-value problem.  For this case, numerical investigation of 
the weak-field Riemannian boundary-value problem was begun in [18], 
and has recently been extended towards the strong-field region [19].  
For weak scalar boundary data, global quantities such as the mass $M$ 
and Euclidean action $I$ appear to scale quadratically, in accordance 
with analytic weak-field estimates [19].  In the limit of strong-field 
scalar boundary data, it may be that a typical pattern will emerge 
numerically for the general 'shape' of the classical Riemannian 
gravitational and scalar fields.  In that case, it might be possible 
to find analytic approximations for the strong-field limit 
(quite different from those valid in the weak-field case), which could 
provide further analytical insight into the solutions of the coupled 
Riemannian boundary-value problem.  In particular, it would be
extremely valuable to have strong-field approximations which were
valid into the complex region, with ${\,}0<\theta <\pi/2{\,}$.  
One might conjecture that, as one approaches the Lorentzian limit 
${\,}\theta\rightarrow 0_{+}{\,}$, for very strong 
spherically-symmetric boundary data, the solutions correspond to 
classical Einstein/scalar solutions which form a singularity, 
surrounded by a black hole.

Feynman's $+i\epsilon$ proposal [13]  for computing quantum amplitudes 
corresponds, in our description, to calculating an amplitude 
(see below) for a complex time-separation 
${\mid}T{\mid}\exp(-i\theta)$, with ${\,}0<\theta\leq\pi/2{\,}$, 
and then taking the limit of the amplitude as 
$\theta\rightarrow 0_{+}{\,}$.

In the case of quantum amplitudes for Lagrangians with Einstein
gravity coupled to matter, with {\it anisotropic} boundary data posed 
in 'field language', such as the case $(h_{ij},\phi)_{I,F}$ above, then 
at least in the asymptotically-flat case with time-interval $T$, one is
inevitably led to consider the complex boundary-value problem, with
${\,}T\rightarrow{\mid}T{\mid}\exp(-i\theta){\,}$, but with
$(h_{ij},\phi)_{I,F}$  unchanged.  Even for fairly small 
${\,}\theta{\,}$, solution of this boundary-value problem is expected 
to smooth out variations or oscillations of the boundary data, 
when one moves into the interior by a few multiples of the relevant 
wavelength.  If the problem is genuinely strongly elliptic, up to gauge,
then one will be able to extend the classical solution analytically 
into the complex.

This and the following papers will use this construction to study, in
particular, a model concerning nearly-spherical collapse to a black
hole, again (in the first instance) with Einstein gravity coupled to a
real massless scalar field $\phi{\,}$, except that, for quantum
reasons (see Sec.2), we consider the simplest locally-supersymmetric
model which contains the bosonic Einstein/matter theory.  For this
$N=1$ supergravity-plus-supermatter model [20], the scalar field $\phi$
becomes complex, with a massless spin-${{1}\over{2}}$ partner, and the 
graviton acquires a spin-${{3}\over{2}}$ gravitino partner.  
It is assumed that there is a 'background' Riemannian 
spherically-symmetric bosonic solution
$(\gamma_{\mu\nu}{\,},{\,}\Phi)$ to the Einstein/scalar 
boundary-value problem, where $\gamma_{\mu\nu}$ is the background 
4-metric and $\Phi$ the background scalar field (taken to be real).  
For simplicity, one can assume that, near the initial surface 
$\Sigma_I{\,}$, the gravitational and scalar fields vary extremely 
slowly, corresponding to diffuse bosonic matter near $\Sigma_I{\,}$.  
For further simplification in the quantum calculation, one can assume 
in the anisotropic case that the spatial restriction $g_{ij}$ of the
exact metric $g_{\mu\nu}{\,}$, together with the scalar field 
$\phi{\,}$, is nevertheless spherically symmetric on the initial
surface $\Sigma_{I}{\;}{\,}(t=0){\,}$ ('no incoming particles at early
retarded times'), although anisotropic on the final surface 
$\Sigma_F{\,}$.  In calculating the Lorentzian quantum amplitude,
one would like to take $\Sigma_{F}$ at a sufficiently late time $T$ 
that all the quantum radiation due to the evaporation of the black
hole will by then have been emitted.  In the Riemannian or complex 
version, this corresponds to choosing final data $(h_{ij},\phi)_F$ 
which are nearly spherically symmetric, allowing only for a
distribution of weak-field (anisotropic) graviton and scalar data on
that part of $\Sigma_F$ (say, $R_{0}<r<R_{1}{\,}$, for some large
radii $R_{0}{\,},R_{1}$), which corresponds to the arrival of radiation 
at $\Sigma_F{\,}$.  The classical back-reaction of the radiation on
the geometry can be described as follows: on $\Sigma_F{\,}$, there is 
a slowly-varying mass function $m(r)$, where $r$ is an intrinsic
radial coordinate.  The radial rate of change $m'(r)$ is given 
{\it via} the Einstein equations, in terms of the averaged energy 
output in radiation [21,22].  Here, $m(r)$ will be extremely small in 
a region around $r=0{\,}$, will then increase very slowly through the 
radiation region, and will then settle at $M_I$ for large $r{\,}$, where 
$M_I$ denotes the conserved ADM (Arnowitt-Deser-Misner) mass of the 
space-time [23,24], as measured, say, at the initial surface 
$\Sigma_I{\,}$.

In the locally-supersymmetric models of $N=1$ supergravity coupled to
supermatter, such as those in [20,25], the quantum amplitude for a
particular configuration of bosonic final data, posed on the surface 
$\Sigma_F{\,}$, is expected to be proportional to $\exp(-I)$ or 
$\exp(iS){\,}$, possibly including finite loop corrections (see Sec.2) 
[15,26,27].  Here $I$ and $S$ are, respectively, the Euclidean and 
Lorentzian classical action corresponding to the solution of the above 
boundary-value problem, with given spherically-symmetric initial data 
(say) on $\Sigma_I{\,}$.  Non-zero fermionic boundary data and
resulting classical actions can also be included, provided that the 
boundary data are suitably posed [28-30].  The resulting classical 
fermionic fields and action will then be elements of a Grassmann
algebra, as usual in the holomorphic representation for fermions 
[31].  (In Lorentzian signature, this point of view has also been
taken in [32], concerning the Cauchy problem for $N=1$ supergravity.)  
Of course, for this statement to make sense, one must again write 
${\,}T{\,}={\,}{\mid}T{\mid}\exp(-i\theta)$ and take the limit as 
$\theta\rightarrow 0_{+}{\,}$.  As follows from Sec.5 and is described 
in detail for scalar boundary data in Paper II, this gives a Gaussian 
form for the weak-field amplitude.

The above description refers to the general unitary evolution of
states in the quantum field theory, assuming that the theory does
indeed have meaningful quantum amplitudes -- this latter point is
discussed further in Sec.2, particularly in relation to local
supersymmetry.  Thus, from this viewpoint, there is no question of
loss of quantum coherence or of information.  From 1976 until 
July 2004, the most generally accepted option for the end-point 
of gravitational collapse to a black hole was, in fact, that quantum 
coherence or information would be lost [2,6].  The somewhat abrupt 
change in opinion since then [10] now makes it possible at last to begin
publishing our late-1990's work on an alternative option, outlined
above in this Introduction, namely, that there are quantum amplitudes
(not just probabilities) for final outcomes, and that the end-state is
a combination of outgoing radiation states.  Two letters describing 
the basic outline of this work have appeared or will shortly 
appear [11,33].

The present paper is concerned with setting up some of the underlying
language and results that will be needed for further applications (see
below).  In Sec.2, as mentioned above, we discuss the conditions under 
which amplitudes in quantum field theory (incorporating Einstein
gravity) are meaningful: almost certainly, this restricts one to 
theories invariant under local supersymmetry, that is, to supergravity 
or to suitable models of supergravity with supermatter.  It is
expected that such models give finite amplitudes, in the present 
description [15,26,27].  In Sec.3, we consider the separation of the 
gravitational and scalar fields into spherically-symmetric background 
parts $(\gamma_{\mu\nu}{\,},{\,}\Phi)$ as above, and non-spherical 
perturbations at linear and higher orders.  The classical field
equations are considered.  In particular, the (classical)
back-reaction is considered, in which terms up to quadratic in the
perturbed metric and scalar field provide an extra effective
energy-momentum source for the gravitational field.  High-frequency
averaging is also described; this smoothes out the 'random' effects of
the emitted black-hole radiation, to give a simpler treatment of the
coupled gravity-plus-radiation system.  Sec.4 is concerned with the 
decomposition of the perturbed part of the scalar field into angular 
harmonics.  Following this  decomposition, the resulting 
$(\tau{\,},r)$-dependent classical scalar field equation is considered;  
this will be essential in the calculation of scalar-field amplitudes 
in Paper II.  In Sec.5, we study the action of classical solutions 
of the Einstein/massless-scalar field equations; this again will be 
needed in Paper II.  Sec.6 contains the Conclusion.

Here, we also describe briefly the main content of this project as
presently envisaged.  Paper II is concerned with the calculation 
of the quantum amplitude for perturbations on the final surface 
$\Sigma_F$ which are purely spin-$0{\,}$, whereas, for simplicity 
in that calculation, the gravitational field on $\Sigma_F$ is taken 
to be exactly spherically symmetric.  Next, we will relate the 
description of Paper II to the Bogoliubov transformations [33] which 
are familiar from many earlier treatments of particle creation [3,34,35].  
In the course of this, the relation between the Vaidya metric [21] 
and the space-time geometry in the region containing the outgoing 
radiation flux will be worked out carefully.  The quantum amplitudes 
or wave functionals for our 'complex-rotated-$T$' problem, as in 
Paper II, are further related to an alternative description in terms 
of coherent and squeezed states; this points to a more universal 
validity of the procedure adopted in this project, with further 
applications to quantum amplitudes for inhomogeneities in cosmology, 
for example.  Finally, the procedure of Paper II for spin-0 
perturbations will be generalised to the case of spin-1 Maxwell 
(or Yang-Mills) perturbations, and to spin-2 graviton perturbations.  
The fermionic case $s={1\over 2}$ of massless neutrino perturbations 
will be treated similarly.  The remaining fermionic case of the 
spin-${3\over 2}$ field, needed for locally supersymmetric models, 
is in preparation.  In addition, substantial work has been done 
concerning computer solution of the elliptic or the complexified
version of the spherically-symmetric Einstein/massless-scalar field 
equations, following [18].  Results of this further work should 
soon be presented [19].
\end{section}

\begin{section}{The quantum amplitude for bosonic boundary data}
Consider the 'Euclidean' quantum amplitude to go between prescribed
initial and final purely bosonic data, each given on a 3-surface 
which is 'topologically' (diffeomorphically) $\Bbb R^{3}{\,}$, and each 
carrying an asymptotically-flat 3-metric.  It is further necessary 
to specify the proper (Euclidean) distance $\tau{\,}$, measured 
orthogonally between the two surfaces at spatial infinity.  This 
amplitude will be given (formally, at least) by a Feynman path
integral over all Riemannian infilling 4-geometries together with any 
other bosonic fields, each such configuration being weighted by 
$\exp(-I){\,}$, where $I$ is the 'Euclidean action' of the configuration.  
If this definition is meaningful, one expects that the resulting 
'Euclidean' quantum amplitude has the semi-classical form
$${\rm Amp}{\;}{\,}\sim{\;}{\,}(A_{0}+{\hbar}A_{1}+{\hbar}^{2}A_{2}
+{\ldots}{\;}){\;}{\,}\exp(-I_{B}/{\hbar}){\quad},\eqno(2.1)$$
\noindent
asymptotically in the limit that 
${\,}\bigl(I_{B}/\hbar\bigr)\rightarrow 0{\;}.$
Here, $I_{B}$ is the classical 'Euclidean action' of a Riemannian
solution of the coupled Einstein and bosonic-matter classical field
equations, subject to the boundary conditions.  For simplicity, we
assume that there is a unique classical solution, up to gauge and
coordinate transformations.  But it is quite feasible, in certain
theories and for certain boundary data, to have instead (say) a
complex-conjugate pair of classical solutions [36].

The classical action $I_{B}$ and loop terms $A_{0}{\,},A_{1}{\,},
A_{2}{\,},\ldots{\;}$ depend in principle on the boundary data.  
In the case of matter coupled to Einstein gravity, each of 
${\,}I_{B}{\,},A_{0}{\,},A_{1}{\,},\ldots{\;}$ will also obey 
differential constraints connected with the local coordinate
invariance of the theory, and with any other local invariances such as 
gauge invariance (if appropriate) [15,37].

In particular, when the theory is also invariant under local
supersymmetry, the semi-classical expansion (2.1) may become extremely 
simple [15,26,27].  For example, for $N=1$ supergravity, one has [15]:
$${\rm Amp}{\;}{\,}\sim{\;}{\,}A_{0}{\;}\exp\bigl(-I/{\hbar}\bigr){\quad}.
\eqno (2.2)$$
\noindent
In this theory, the one-loop factor $A_{0}$ is in fact a constant.  
When one allows the boundary data to include both bosonic and
fermionic parts, suitably posed, one expects that a classical 
solution of the coupled bosonic/fermionic field equations will still 
exist.  In this case, $I$ in Eq.(2.2) denotes the classical action, 
including now both bosonic and fermionic contributions.  Related 
properties hold for $N=1$ supergravity coupled to gauge-invariant 
supermatter [25,26].  There will also, for example, be analogous
consequences for locally-supersymmetric models of $N=1$ supergravity 
coupled to supermatter [26,27], which in the simplest case include a 
complex scalar field and a spin-$1\over 2$ field [20]; one expects the 
amplitude semi-classically to be of the form (2.1), with finite loop 
terms $A_{0}{\,},A_{1}{\,},{\ldots}{\;}$ [15,26,27].

In the case (2.2) of $N = 1$ supergravity, the classical action is
all that is needed for the quantum computation.  A corresponding
situation arises with ultra-high-energy collisions, whether between
black holes [38], in particle scattering [39], or in string theory
[40].

In the asymptotically-flat, spatially-$\Bbb R^3$ context appropriate
here, the purely Riemannian case above corresponds, in 
Lorentzian-time language, to a (Lorentzian) time-separation at spatial 
infinity of the usual rotated form ${\,}T={\,}-i\tau{\,}$, where 
$\tau$ is the (positive) imaginary-time separation defined above.  
If the four-dimensional classical bosonic part of the solution 
is to be real, then certainly the bosonic boundary data should be chosen 
to be real.  Following the standard route, one should then study 
the (now complex) bosonic amplitude (2.1) or (2.2), as $T$ is rotated 
through a range of angles ${\,}\theta{\,}$, starting from 
${\,}\theta =\pi/2$ and ending at 
${\,}\theta{\,}=+{\,}\epsilon{\quad}(>0){\,}$, with
$$T{\;}{\,}={\;}{\,}\tau{\;}\exp(-i\theta){\quad}.\eqno (2.3)$$
\noindent
Provided that there continues to exist a (complex) bosonic classical 
solution to the Dirichlet problem, as one rotates ${\,}\theta{\,}$ 
from ${\,}\pi/2{\,}$ towards zero, the expression (2.1) or (2.2) should 
continue to give the form of the quantum amplitude, which should be 
analytic in $\theta$ (among other variables).  In particular, this
would occur if strong ellipticity [16] held for the coupled 
Einstein/bosonic-matter field equations, up to gauge.
\end{section}

\begin{section}{The approximate 4-dimensional metric}

The classical background bosonic fields are at present taken, for
simplicity, to be only the metric $g_{\mu \nu}$ and the massless 
scalar field $\phi{\,}$.  In later work, we shall study cases in which
other-spin fields, with $s={{1}\over{2}}{\,},1$ or 
${\,}{{3}\over{2}}{\,}$, are included as perturbations of a
spherically-symmetric background solution [30,41,42].  The classical 
solutions $(g_{\mu\nu}{\,},\phi)$ of the coupled Einstein/scalar field 
equations below are taken to have a 'background' time-dependent 
spherically-symmetric part $(\gamma_{\mu\nu}{\,},\Phi){\,}$, together 
with a 'small' perturbative part $(h_{\mu\nu}{\,},\phi_{pert})$.  
The perturbative fields $h_{\mu\nu}$ and $\phi_{pert}{\,}$, which live 
on the spherically-symmetric background 4-geometry $\gamma_{\mu\nu}{\,}$, 
can, as usual, be expanded out in terms of sums over tensor (spin-2), 
vector (spin-1) and scalar harmonics [43,44].  Each harmonic is
weighted by a function of the Riemannian time- and radial 
coordinates $(\tau,r){\,}$.

The Einstein field equations are taken in the form
$$G_{\mu\nu}{\;}{\,}
\equiv{\;}{\,}R_{\mu\nu}{\,}-{\,}{1\over 2}{\,}R{\,}g_{\mu\nu}{\;}{\,}
={\;}{\,}8\pi{\,}T_{\mu\nu}{\quad},\eqno (3.1)$$
where $R_{\mu\nu}$ denotes the Ricci tensor, $R$ the Ricci scalar 
and $T_{\mu\nu}$ the energy-momentum tensor.  For a (real) massless
scalar field $\phi{\,}$, one has 
$$T_{\mu\nu}{\;}{\,} 
={\;}{\,}\phi,_{\mu}\phi,_{\nu}
-{\,}{1\over 2}{\,}g_{\mu\nu}{\,}
\bigl(\phi,_{\alpha}\phi,_{\beta}g^{\alpha\beta}\bigr){\quad}.\eqno (3.2)$$  
The gravitational field equations further imply the scalar field
equation (the Laplace-Beltrami equation [45]):
$$\partial_{\mu}\bigl(g^{1/2}{\,}g^{\mu\nu}{\,}\phi,_{\nu}\bigr){\;}{\,} 
={\;}{\,}0{\quad},\eqno (3.3)$$
\noindent
where ${\,}g{\,}$ denotes $\det(g_{\mu\nu}){\,}$, and at present we assume
that $g_{\mu\nu}$ is real Riemannian, whence $g{\,}>{\,}0{\,}$.

The corresponding Riemannian variational principle refers to an action
functional of the form [15]
$$I{\;}{\,}
={\;}{\,}{{-1}\over{16\pi}}{\,}\int d^{4}x{\;}{\,}g^{{1/2}}{\,}R{\;}
+{\;}{1\over 2}{\,}\int d^{4}x{\;}{\,}g^{1/2}{\,}(\nabla\phi)^{2}{\,}
+{\;}{\,}{\rm boundary{\,}contributions}{\quad}.\eqno (3.4)$$    
\noindent  
The appropriate boundary terms will be discussed below in Sec.5.

In the Riemannian case [18], the 'large' or 'background' 4-metric can
be put in the form:
$$ds^2{\;}{\,}
={\;}{\,}e^{b}{\,}d{\tau}^{2}+{\,}e^{a}dr^{2} 
+r^{2}(d{\theta}^{2}+{\,}{\sin}^{2}{\theta}{\;}d{\phi}^2){\quad},
\eqno (3.5)$$
\noindent
where
$$b{\;}{\,}={\;}{\,}b(\tau,r){\;}{\,},
{\qquad}{\qquad}a{\;}{\,}={\;}{\,}a(\tau,r){\;}{\,}.\eqno (3.6)$$
\noindent
If the gravitational field were exactly spherically symmetric, as in
Eq.(3.5), and if the scalar field were also rotationally invariant,
being of the form ${\,}\phi(\tau, r)$, then the Riemannian 
spherically-symmetric scalar and Einstein field equations would 
hold [18].  The scalar field equation reads:
$$\ddot\phi{\;}+{\;}e^{{b-a}}{\;}{\phi}^{\prime\prime}{\;}
+{\;}{1\over 2}{\,}\bigl(\dot a -\dot b\bigr){\;}\dot{\phi}{\;} 
+{\;}r^{-1}{\;}e^{{b-a}}{\;}\bigl(1{\;}+{\;}e^{a}\bigr){\;}\phi'{\;}{\,}
={\;}{\,}0{\quad},\eqno (3.7)$$
\noindent
where $(\dot{\;\,})$ denotes ${\partial}(~)/{\partial}{\tau}{\,}$ 
and $(~)^{\prime}$ denotes ${\partial}(~)/{\partial}r{\,}$.  
Together with Eq.(3.7), a slightly redundant set of gravitational 
field equations is given by:
$$\eqalignno{a^\prime{\;}{\,}&={\;}{\,}-4\pi{\;}r{\;}
\bigl(e^{a-b}{\;}{\dot\phi}^{2}{\;}
-{\;}{\phi^{\prime}}^2{\,}\bigr){\;}
+{\;}r^{-1}{\;}\bigl(1-e^a\bigr){\;},&(3.8)\cr
b^{\prime}{\;}{\,}&={\;}{\,}-4\pi{\;}r{\;}
\bigl(e^{a-b}{\;}{\dot\phi}^{2}{\;}
-{\;}{\phi^{\prime}}^2\bigr){\;}
-{\;}r^{-1}{\;}\bigl(1{\;}-{\;}e^a\bigr){\;},&(3.9)\cr
\dot{a}{\;}{\,}&={\;}{\,}8\pi{\;}r{\;}\dot{\phi}{\;}\phi^{\prime}
{\;},&(3.10)\cr}$$
$$\ddot a +e^{b-a}{\,}b^{\prime\prime}
+{1\over 2}\bigl(\dot a -\dot b\bigr)\dot a{\,} 
-{\,}r^{-1}e^{b-a}\bigl(1-e^a\bigr)\bigl(b^{\prime}+2r^{-1}\bigr){\;}{\,}
={\;}{\,}8\pi{\,}\bigl({\dot\phi}^2
+e^{b-a}{\,}{\phi^\prime}^2\bigr){\;}.\eqno(3.11)$$ 
\smallskip
\indent
The metric and the classical field equations in Lorentzian 
signature [17], or for certain types of complex metrics, can be
derived from the above by the formal replacement
$$t{\;}{\,}={\;}{\,}\tau{\;}e^{-i\theta}{\;},\eqno(3.12)$$
\noindent 
where $\theta$ is independent of 4-dimensional position, and where
$\theta$ should be rotated from $0$ to $\pi/2{\,}$.

In the bosonic black-hole evaporation problem, the classical Riemannian
metric and scalar field will not be exactly spherically symmetric;
similarly for any non-zero spin-$1\over2$ and spin-$3\over2$
classical (odd Grassmann-algebra-valued [15]) fermionic solutions in the 
locally-supersymmetric generalisation [20].  In particle language,
rather than the field language which is mostly being used in this
paper, huge numbers of gravitons and scalar particles will continually 
be given off by the black hole (together with any fermions allowed by the
model), leading to an effectively stochastic distribution, in which,
for any given spin $s{\,}$, the field fluctuates around a
spherically-symmetric reference field.

Consider, for example, gravitational and scalar perturbations about a
Riemannian spherically-symmetric reference 4-metric $\gamma_{\mu\nu}$
and reference scalar field $\Phi{\,}$.  In perturbation theory in general
relativity [46], one considers a one-parameter (or many-parameter)
family of 4-metrics, here of the form
$$g_{\mu\nu}(x,\epsilon){\;}{\,}
={\;}{\,}\gamma_{\mu\nu}(x){\;}
+{\;}\epsilon{\;}h^{(1)}_{\mu\nu}(x){\;}
+{\;}\epsilon^2{\;}h^{(2)}_{\mu\nu}(x){\;}+\ldots{\quad},\eqno (3.13)$$
\noindent
where $h^{(1)}_{\mu\nu}$ is the first-order metric perturbation,
$h^{(2)}_{\mu\nu} $ is the second-order perturbation, etc.
Throughout, the superscript $^{(0)}$ will refer to the background,
while $^{(1)}$ denotes terms linear in perturbations, etc.  Indices
are to be raised and lowered using the background metric 
$\gamma^{\mu\nu},\gamma_{\mu\nu}{\,}.$  Covariant derivatives 
with respect to the background geometry are denoted either 
by a semi-colon ${~}_{;\alpha}$ or equivalently by 
$\nabla_\alpha{\,}$.

Analogously, we split a real massless scalar field $\phi$ into a
spherically-symmetric background piece $\Phi(\tau,r)$ and a
(non-spherical) perturbation:
$$\phi(x,\epsilon){\;}{\,}
={\;}{\,}\Phi(\tau,r){\;}+{\;}\epsilon{\;}\phi^{(1)}(x){\;}
+{\;}{\epsilon}^{2}{\;}\phi^{(2)}(x){\;}
+\ldots{\quad}.\eqno (3.14)$$
\noindent
The spherically-symmetric background part $\Phi$ will be non-zero if
the background scalar data $\phi$ at early and late Euclidean times
$\tau$ contain a non-trivial spherically-symmetric component.  The
perturbation fields ${\,}\phi^{(1)}(x),{\;}\phi^{(2)}(x),{\;}\ldots{\,}$ 
will, in general, contain all non-spherical angular harmonics.  
These fields must be chosen such that the entire coupled Einstein/scalar 
system satisfies the classical field equations, as well as agreeing 
with the prescribed small non-spherical perturbations in the initial 
and final data, both gravitational and scalar.  The effective 
energy-momentum source for the spherically-symmetric part 
$\gamma_{\mu\nu}$ of the metric includes contributions formed 
quadratically from the non-spherical gravitational and scalar parts 
$h^{(1)}_{\mu\nu}$ and $\phi^{(1)}{\,}$ --- see Eqs.(3.27-29) below 
for further detail.

In the simplest case, one can restrict attention to the exactly
spherically-symmetric Riemannian model of Eqs.(3.5-11).  Then the
background metric $\gamma_{\mu\nu}$ will correspond to the metric
(3.5,6), with respect to suitable coordinates, and $\Phi(\tau,r)$
here will correspond to the $\phi(\tau, r)$ of those equations.  
This Riemannian boundary-value problem, involving a system of coupled partial
differential equations in two variables $(\tau,r)$, has been studied
numerically in [18] for particular choices of boundary data, and is
currently being investigated in much greater detail [19].

By contrast, the Lorentzian-signature version of the
spherically-symmetric classical Einstein/scalar system must be studied
as an initial-value evolution problem, in order to be 
well-posed [17].  One conceivable initial profile for the scalar
field, which has been much studied in the Lorentzian-signature numerical 
problem [47,48], is an ingoing 'Gaussian' shell of scalar radiation.  
To define such initial data, work in a nearly-flat space-time at very 
early times (large negative Lorentzian time-coordinate $t$).  
Define an advanced null coordinate
$$v{\;}{\,}={\;}{\,}t{\,}+{\,}r{\quad}.\eqno (3.15)$$
\noindent
The incoming 'Gaussian' shell is asymptotically, at early times, 
$$\Phi(t,r){\;}{\,}
\sim{\quad}{{f_{0}{\;}v^{k+1}}\over{r}}{\;}{\,}
\exp\Biggl[-{\biggl(}{{v-v_{0}}\over {\Delta}}{\biggr)}^d{\,}
{\Biggr ]}{\quad},\eqno (3.16)$$     
\noindent
where $f_0{\,},k{\,},d{\,},r_0{\,}$ and $\Delta$ are all positive 
real parameters.  Here, the radial extent, $L_0{\,}$, of the 
'Gaussian' is given by  ${\,}L_0{\,}\sim{\,}\Delta{\,}$. The numerical 
evolution of such initial data provides a model of spherical collapse.  
In particular, two main qualitatively different ${\it r\acute egimes}$ 
of initial data can be distinguished.  Firstly, if $L_0$ or $\Delta$ 
is too large, then the initial data are 'diffuse', there is little 
self-interaction, and the incoming scalar profiles pass more or less 
straight through each other, leaving behind nearly-flat space-time 
plus small perturbations.  Secondly, if $\Delta$ (or $L_0$) is less 
than a certain critical value, the interaction is sufficiently
non-linear that a black hole forms.

Returning to the Riemannian or to the complex case, one can expand out
the Einstein field equations (3.1,2) in powers of $\epsilon{\,}$.
At lowest order $(\epsilon^0)$, one has the background Einstein 
and scalar field equations 
$$\eqalignno{R^{(0)}_{\mu\nu}{\,}
-{\,}{1\over 2}{\,}R^{(0)}{\,}\gamma_{\mu\nu}{\;}{\,}&
={\;}{\,}8\pi{\,}T^{(0)}_{\mu\nu}{\quad}, &(3.17)\cr
\gamma^{\mu\nu}{\,}\Phi_{;\mu\nu}{\;}{\,}&
={\;}{\,}0{\quad}, &(3.18)\cr}$$
where $R^{(0)}_{\mu\nu}$ denotes the Ricci tensor and $R^{(0)}$
denotes the Ricci scalar of the background geometry
$\gamma_{\mu\nu}{\,}$, and where a semi-colon now denotes covariant
differentiation with respect to the background geometry.  Further,
$$T^{(0)}_{\mu\nu}{\;}{\,}
={\;}{\,}\Phi,_{\mu}{\,}\Phi,_{\nu}{\,}
-{\,}{1\over 2}{\,}\gamma_{\mu\nu}{\,}
\bigl(\Phi,_{\alpha}\Phi,_{\beta}{\,}\gamma^{\alpha\beta}\bigr)
\eqno (3.19)$$
\noindent
denotes the background spherically-symmetric energy-momentum tensor.
These field equations are equivalent to Eqs.(3.7-11), when coordinates
are taken as in Eqs.(3.5,6).

The linearised $(\epsilon^1)$ part of the Einstein equations reads
(see Section 35.13 of [24])
$$\eqalign{\bar{h}^{(1)}_{\mu\nu;\sigma}{}^{;\sigma}  
&-2{\,}\bar{h}^{(1)}_{\sigma(\mu}{}^{;\sigma}{}_{;\nu)} 
-2{\,}R^{(0)}_{\sigma\mu\nu\alpha}{\,}\bar{h}^{(1)\sigma\alpha} 
-2{\,}R^{(0)\alpha}{}_{(\mu}\bar{h}^{(1)}_{\nu)\alpha}\cr 
+{\,}&\gamma_{\mu\nu}{\,}\bigl(\bar{h}^{(1)}_{\alpha\beta}{}^{;\alpha\beta}
-\bar{h}^{(1)\alpha\beta}R^{(0)}_{\alpha\beta}{\,}\bigr)
+{\,}\bar{h}^{(1)}_{\mu\nu}{\,}R^{(0)}{\;}{\,}
={\;}{\,}-16\pi{\,}T^{(1)}_{\mu\nu}{\quad}.\cr}
\eqno(3.20)$$
\noindent
Here, $\bar{h}^{(1)}_{\mu\nu}$ is defined by
$$\bar{h}^{(1)}_{\mu\nu}{\;}{\,}
={\;}{\,}h^{(1)}_{\mu\nu}-{1\over 2}{\,}\gamma_{\mu\nu}{\,}h^{(1)}{\quad},
\eqno(3.21)$$
\noindent
where 
$$h^{(1)}{\;}{\,}={\;}{\,}h^{(1)\mu}_{\mu}{\quad}.\eqno(3.22)$$
\noindent
Here, $R^{(0)}_{\sigma\mu\nu\alpha}$ denotes the Riemann tensor of the
background geometry $\gamma_{\mu\nu}{\,}$.  Further, $T^{(1)}_{\mu\nu}$
denotes the linearisation or $O(\epsilon^1)$ part of the
energy-momentum tensor $T_{\mu\nu}(x,\epsilon)$.  Explicitly,
$$T^{(1)}_{\mu\nu}{\;}{\,}
={\;}{\,}2{\,}\nabla_{(\mu}\phi^{(1)}{\,}\nabla_{\nu)}\Phi{\,}
-{\,}\gamma_{\mu\nu}\bigl(\nabla_{\alpha}\Phi\bigr)
\bigl(\nabla^{\alpha}\phi^{(1)}\bigr){\,}
+{\,}{1\over 2}{\,}\bigl(\gamma_{\mu\nu}{\,}h^{(1)\sigma\rho}
-h^{(1)}_{\mu\nu}{\,}\gamma^{\sigma\rho}\bigr)
\bigl(\nabla_{\sigma}\Phi\bigr)\bigl(\nabla_{\rho}\Phi\bigr).\eqno (3.23)$$
\smallskip
\indent
The linearised Einstein equations (3.20-23) are most easily studied 
in a 'linearised harmonic gauge' [24,49], in which, by an infinitesimal
coordinate transformation, one arranges that
$$\bar{h}^{(1)~;\alpha}_{\mu\alpha}{\;}{\,}={\;}{\,}0 \eqno (3.24)$$
\noindent
everywhere.  Since the gravitational background $\gamma_{\mu\nu}$ is
spherically symmetric, the linearised Einstein equations (3.20-23) can 
be further decomposed into three independent sets of equations.  These
describe repectively scalar (spin-0) perturbations associated with
matter-density changes $\bigl(T^{(1)}_{\tau\tau}\bigr)$, vector (spin-1)
perturbations associated with matter-velocity changes 
$\bigl(T^{(1)}_{\tau i}\bigr)$, and gravitational radiation (spin-2) 
associated with anisotropic stresses $\bigl(T^{(1)}_{ij}\bigr)$ [50].

The linearised $(\epsilon^{1})$ part of the scalar field equation
(3.3) yields
$$\gamma^{\mu\nu}{\,}\phi^{(1)}_{;\mu\nu}{\,}
-\bigl(\bar{h}^{(1)\mu\nu}{\,}\Phi,_{\nu}\bigr)_{;\mu}{\;}{\,} 
={\;}{\,}0{\quad}.\eqno (3.25)$$
\noindent
Thus, the linearised Einstein and linearised scalar field equations 
(3.20-23,25) are coupled.

At order $\epsilon^{2}{\,}$, the gravitational field equations give the
second-order contribution $G^{(2)}_{\mu\nu}$ to the Einstein tensor
$$G_{\mu\nu}{\;}{\,}
={\;}{\,}R_{\mu\nu}{\,}-{\,}{1\over 2}{\,}R{\,}g_{\mu\nu}{\quad};
\eqno (3.26)$$
\noindent 
details are given below. We note that $G^{(2)}_{\mu\nu}$ includes a
well-known contribution quadratic in the first-order perturbations
$h^{(1)}_{\mu\nu}$ and their derivatives --- see Eq.(35.58b) of [24].
This part represents an effective energy-momentum-stress density
due to the gravitational perturbations, including gravitons.
$G^{(2)}_{\mu\nu}$ also contains contributions at quadratic order,
formed from the background $\Phi$ and the  linearised $\phi^{(1)}$,
together with $\gamma_{\mu\nu}$ and $h^{(1)}_{\mu\nu}{\,}$.  These parts
represent the $O(\epsilon^{2})$ contribution of the scalar-field
energy-momentum tensor $T_{\mu\nu}$ of Eq.(3.2).

Explicitly, one finds, after a lengthy calculation [9], that the 
Einstein equations, to quadratic order in perturbations, read
$$G^{(0)}_{\mu\nu}{\;}{\,} 
={\;}{\,}8\pi{\,}T^{(0)}_{\mu\nu}{\,}+{\,}8\pi{\,}T^{(2)}_{\mu\nu}{\,} 
+{\,}8\pi{\,}T^{\prime}_{\mu\nu}{\,}-{\,}G^{(1)}_{\mu\nu}{\quad}. 
\eqno (3.27)$$
\noindent
Here,
$$\eqalign{T^{(2)}_{\mu\nu}{\;}{\,} 
={\;}{\,}\nabla_{\mu}\phi^{(1)}{\,}&\nabla_{\nu}\phi^{(1)}
-{1\over 2}{\,}\gamma_{\mu\nu}{\,}\gamma^{\rho\sigma}{\,}
{\,}\nabla_{\rho}\phi^{(1)}{\,}\nabla_{\sigma}\phi^{(1)}
+\Bigl(\gamma_{\mu\nu}{\,}h^{(1)\sigma\rho}
-h^{(1)}_{\mu\nu}{\,}\gamma^{\sigma\rho}\Bigr)
\nabla_{\sigma}\Phi{\;}\nabla_{\rho}\phi^{(1)} \cr
&+{1\over 2}\Bigl(h^{(1)}_{\mu\nu}{\,}h^{(1)\sigma\rho}
-\gamma_{\mu\nu}{\,}h^{(1)\sigma\alpha}{\,}h^{(1)}_{\alpha}{}^{\rho}\Bigr)
\nabla_{\sigma}\Phi{\;}\nabla_{\rho}\Phi\cr}\eqno (3.28)$$ 
\noindent
and
$$\eqalign{8\pi{\,}T'_{\mu\nu}{\;}{\,}
&={\;}{\,}{1\over 4}{\,}
\Bigl(\bar{h}^{(1)\sigma\rho}{}_{;\mu}{\;}h^{(1)}_{\sigma\rho}{}_{;\nu}{\,}
-{\,}2{\,}\bar{h}^{(1)}_{\alpha\sigma}{}^{;\alpha}{\;}
\bar{h}^{(1)\sigma}{}_{(\mu;\nu)}\Bigr)
-{1 \over 2}{\,}\bar{h}^{(1)\sigma}{}_{(\mu}R^{(0)}_{\nu)
\rho\sigma\alpha}\bar{h}^{(1)\alpha\rho}\cr
&+{1 \over 2}{\,}\bar{h}^{(1)}_{\sigma(\mu}{\,}
R^{(0)}_{\nu)\alpha}{\,}\bar{h}^{(1)\alpha\sigma}
-{1 \over 2}{\,}h^{(1)\sigma}{}_{(\mu}\bar{h}^{(1)}_{\nu)\sigma}{\,}R^{(0)}
-{\,}8\pi{\,}T_{\sigma(\mu}^{(1)}\bar{h}^{(1)}_{\nu)}{}^{\sigma}\cr
&-4\pi{\,}\gamma_{\mu\nu}{\,}
\Bigl(2{\,}\bar{h}^{(1)\sigma\rho}{\,}
\nabla_{\sigma}\phi^{(1)}{\,}\nabla_{\rho}\Phi{\,}
+{\,}\phi^{(1)}{\,}\nabla_{\sigma}\nabla^{\sigma}\phi^{(1)}
-{\,}\bar{h}^{(1)\sigma\rho}{\,}h^{(1)}_{\sigma}{}^{\beta}{\,}
\nabla_{\rho}\Phi{\,}\nabla_{\beta}\Phi\Bigr)\cr
&+{\,}C^{\sigma}_{\mu\nu ;\sigma}{\quad},\cr}\eqno (3.29)$$
\noindent
where the explicit form of $C^{\sigma}_{\mu\nu}$ will not be used
here.  The above expressions will be needed particularly in studying 
the Vaidya metric [21], which describes approximately the late-time 
region of the geometry following gravitational collapse to a black
hole, containing a nearly-steady outgoing flux of radiation.  
The Einstein field equations, averaged over small regions, give 
the contribution of massless scalar particles, gravitons, etc., 
to the nearly-isotropic flux.

Physically, for the Riemannian or complex boundary-value problem
discussed in Secs.1 and 2, the $O(\epsilon)$ perturbations in the 4-metric
$g_{\mu\nu}$ and scalar field $\phi$, relative to the
spherically-symmetric background solution $(\gamma_{\mu\nu}{\,},{\,}\Phi)$,
should arise classically from $O(\epsilon)$ perturbations away from
spherical symmetry in the boundary data ${\,}g_{ij}{\,},{\,}\Phi$ 
(or ${\,}\partial\Phi/\partial n$) at the initial and final surface.
Moreover, as mentioned earlier, provided that the perturbed boundary
data contain numerous high harmonics, the 4-dimensional perturbations
in the interior would be expected to have an effectively stochastic
nature.  When averaged over a number of wavelengths, the effective
perturbative energy-momentum tensor above, $T^{EFF}_{\mu\nu}$, will
yield a spherically-symmetric smoothed-out quantity 
$<T^{EFF}_{\mu\nu}>{\,}$ [51,52].  In the locally-supersymmetric version of
this theory, the energy-momentum tensor due to the spin-$1\over2$ and
spin-$3\over2$ fields will also contribute to 
${\,}<T^{EFF}_{\mu\nu}>{\,}$.  This will account, in particular, for 
the gradual loss of mass by radiation of a black hole in the 
nearly-Lorentzian sector (that is, in the case of a time-interval at 
infinity of the form ${\,}T={\,}\tau{\,}\exp(-i\theta){\,}$, 
where $\tau$ is real and positive, and $\,\theta = \epsilon$ is small 
and positive).  Although $<T^{EFF}_{\mu\nu}>{\,}$ is small, being of 
order $\epsilon^2$, its effects on the black-hole geometry, including
the mass, will build up in a secular fashion, over a time-scale of order 
$O(\epsilon^{-2})$.  Such secular behaviour appears often in
perturbation problems [53,54] --- for example, in the familiar treatment
of perihelion precession for nearly-circular orbits in the
Schwarzschild geometry [55].  In our boundary-value problem, whether
regarded as classical or quantum, the initial boundary data will be
spread over a 'background' extent of $O(1)$ in the radial coordinate
$r$ on the initial surface $\Sigma_I{\,}$.  But, corresponding to the
$O(\epsilon^{-2})$ time-scale for the black hole to radiate, the final
data on $\Sigma_F$ will be spread over a radial-coordinate scale of
$O(\epsilon^{2})$.  Thus, even the classical boundary-value problem
here is an example of singular perturbation theory [53,54].

The standard treatment of high-frequency averaging in general
relativity was given by Brill and Hartle [51] and by Isaacson [52].
Let $<\;>$ denote an average over a time $T_0$ much longer than typical
wave periods, together with a spatial average over several wavelengths
$\bar\lambda{\,}$.  Then
$$\eqalignno{<g_{\mu\nu}>{\;}{\,}
&={\;}{\,}\gamma_{\mu\nu}{\quad},{\qquad}{\qquad}<\phi>{\;}{\,}
={\;}{\,}\Phi{\quad},&(3.30)\cr
<\phi^{(1)}>{\;}{\,}&={\;}{\,}0{\quad},
{\qquad}{\qquad}<h^{(1)}_{\mu\nu}>{\;}{\,}={\;}{\,}0{\quad},&(3.31)\cr
<\partial_{\sigma} h^{(1)}_{\mu\nu}>{\;}{\,}&={\;}{\,}0{\quad},
{\qquad}{\qquad}<\partial_{\sigma}\partial_{\rho}h^{(1)}_{\mu\nu}>{\;}{\,} 
={\;}{\,}0{\quad}.&(3.32)\cr}$$
\noindent
Indeed, 
$$<C^{(0)}>{\;}{\,}={\;}{\,}C^{(0)}{\quad},\eqno(3.33)$$
for any background quantity $C^{(0)}$.  Rules for manipulating these
averages in the high-frequency aproximation are set out in [52].
Under integrals, the average of total divergences can be neglected.
For example,
$$<h^{(1)\alpha~~;\beta}_{~~~~\mu}{\;}h^{(1)}_{\beta\nu;\alpha}>{\;}{\,} 
={\;}{\,}-<h^{(1)\alpha~~;\beta}_{~~~~\mu~~~~;\alpha}{\;}
h^{(1)}_{\beta\nu}>{\quad}.\eqno(3.34)$$
\noindent
Further, covariant derivatives commute for high-frequency waves.  The
rules (3.30-34) imply
$$< T^{(1)}_{\mu\nu}>{\;}{\,}={\;}{\,}0{\quad},\eqno(3.35)$$
$$<T^{(2)}_{\mu\nu}>{\;}{\,} 
={\;}{\,}<\nabla_{(\mu}\phi^{(1)}\nabla_{\nu)}\phi^{(1)}
-{1\over 2}\gamma_{\mu\nu}\nabla_{\alpha}\phi^{(1)}{\,}
\nabla^{\alpha}\phi^{(1)}>{\quad}.\eqno(3.36)$$

We can now rewrite the background field equations (3.7-11) in a form
smoothed out by averaging over a number of wavelengths of the scalar
and gravitational perturbations [51,52].  The equation which includes
the quadratic-order contribution of the perturbations as a source for
the background geometry reads:
$$G^{(0)}_{\mu\nu}(\gamma){\;}{\,} 
={\;}{\,}8\pi{\,}T^{(0)}_{\mu\nu} 
+{\,}8\pi{\,}\epsilon^{2}{\,}
\Bigl(<T^{(2)}_{\mu\nu}>+<T^{\prime}_{\mu\nu}>\Bigr){\quad}.\eqno(3.37)$$
\noindent
The terms in this equation vary over length-scales
$>>\bar\lambda{\,}$.  The 'source equation' for $h^{(2)}_{\mu\nu}{\,}$, 
analogous to Eq.(3.20) for $h^{(1)}_{\mu\nu}{\,}$, is
$${\,}G^{(1)}_{\mu\nu}\bigl(\gamma,h^{(2)}\bigr){\;}{\,} 
={\;}{\,}8\pi\Bigl(T^{\prime}_{\mu\nu}{\,}
-<T^{\prime}_{\mu\nu}>\Bigr)
+{\,}8\pi\Bigl(T^{(2)}_{\mu\nu}{\,}-<T^{(2)}_{\mu\nu}>\Bigr){\quad}.
\eqno(3.38)$$
\noindent
Here, the left-hand side ${\,}G^{(1)}_{\mu\nu}(\gamma{\,},h^{(2)}){\,}$ 
denotes the first perturbation of the Einstein tensor $G_{\mu\nu}$ about the
background metric $\gamma_{\mu\nu}{\,}$, but with its linear argument
taken to be $h^{(2)}_{\mu\nu}$ rather than $h^{(1)}_{\mu\nu}{\,}$. 
Thus, $-{\,}2{\,}G^{(1)}_{\mu\nu}(\gamma{\,},h^{(1)})$ is given 
by the left-hand side of Eq.(3.20), subject to Eqs.(3.21,22).  Hence, 
the left-hand side of Eq.(3.38) is linear in $h^{(2)}_{\mu\nu}$ and 
its derivatives, whereas the right-hand side is quadratic in 
first-order fluctuations.  By contrast with Eq.(3.37), the terms in 
Eq.(3.38) vary over length-scales of order $\bar\lambda{\,}$.
\end{section}

\begin{section}{Scalar field: harmonic description}
Consider small bosonic perturbations $\phi^{(1)}$ and 
$h^{(1)}_{\mu\nu}{\,}$, obeying the linearised classical field equations 
(3.20) and (3.25) about a spherically-symmetric classical solution 
$(\Phi{\,},{\,}\gamma_{\mu\nu})$ of the Riemannian field equations 
(3.7-11) for Einstein gravity coupled minimally to a massless scalar
field.  The background spherically-symmetric data for $\Phi$ and 
$\gamma_{\mu\nu}$ are posed, as in Secs.1,2, on the initial and final 
3-dimensional boundaries, separated at spatial infinity by a 
'Euclidean time-separation' $\tau >0{\,}$.  Similarly, the linearised 
classical perturbations $\phi^{(1)}$ and $h^{(1)}_{\mu\nu}$ are to be 
regarded as the solutions to a coupled linear elliptic problem,
subject to prescribed linearised perturbations $\phi^{(1)}$ (say) and 
$h^{(1)}_{ij}$ on the initial and final boundaries.

Because of the spherical symmetry of the background
$(\Phi{\,},\gamma_{\mu\nu})$, one may expand the Riemannian 
4-dimensional perturbation $\phi^{(1)}$ in the form
$$\phi^{(1)}(\tau{\,},r{\,},\theta{\,},\phi){\;}{\,} 
={\;}{\,}{1\over r}{\;}
\sum^{\infty}_{\ell = 0}{\;}\sum^{\ell}_{m = -\ell}Y_{\ell m}(\Omega){\;}
R_{\ell m}(\tau{\,},r){\quad}.\eqno(4.1)$$
\noindent
Here, $Y_{\ell m}(\Omega)$ denotes the $(\ell, m)$ scalar spherical 
harmonic of [56].

Similarly, a generic Riemannian metric pertubation $h^{(1)}_{\mu\nu}$
may be expanded out as a sum over spin-2 (tensor), spin-1 (vector)
and spin-0 (scalar) $(\ell{\,}, m)$ harmonics, each weighted by a function
of $\tau$ and $r$ [43,44,57-60].  The amplitudes for graviton
(spin-2) emission following black-hole collapse will be treated
subsequently, including further details of the spin-2 harmonics.  
But note that, because of the coupled nature of the linearised field 
equations (3.20,25) for $\phi^{(1)}$ and $h^{(1)}_{\mu\nu}{\,}$, the 
resulting linear field equations in $\tau$ and $r$ for 
$R_{\ell m}(\tau, r)$ of Eq.(4.1) and its gravitational analogues will 
also be coupled in the strong-field 'collapse' region of the
`space-time'.

The boundary conditions on the radial functions 
$R_{\ell m}(\tau{\,},r)$ as $r \rightarrow 0$ follow from the
regularity there of the whole Riemannian solution, consisting of 
$\phi$ and the 4-metric $g_{\mu\nu}$ (but viewed in 
'nearly-Cartesian coordinates' near $r=0$). This regularity of the 
solution in turn follows since the coupled field equations are 
'elliptic {\it modulo} gauge'.  For simplicity, the boundary data, 
on both the initial and the final 3-surface, must be chosen to be 
sufficiently regular or smooth over $\Bbb R^3{\,}$, in addition 
to being asymptotically flat.  Even when one takes a complex 
Lorentzian time-separation-at-infinity
$$T{\;}{\,}={\;}{\,}\tau{\,}\exp(-i\theta){\quad},\eqno(4.2)$$
as in Eq.(2.3), with ${\,}0<\theta\leq\pi/2{\,}$, one expects that
the field equations (up to gauge) will be strongly elliptic [16],
whence all
classical fields must be analytic in the interior of the large
cylindrical boundary formed by the initial and final surfaces,
together with a surface at large $r{\,}$.

Suppose that the boundary conditions on the final surface are taken to
consist of weak and very diffuse scalar and gravitational fields,
regarded as perturbations of flat 3-space ${\Bbb E}^3{\,}$. (One also
requires that the ADM (Arnowitt-Deser-Misner) mass of the final
intrinsic boundary 3-metric $g_{ij}{\,}$, computed from the $(1/r)$
part of the fall-off of $g_{ij}$ to the flat metric $\delta_{ij}$ 
[23,24,61] should be the same as the ADM mass of the initial surface.
This will be discussed further in Sec.5 below.)  Physically, such weak
and diffuse final boundary data may be imagined to be a possible
late-time remnant of gravitational collapse, namely a snap-shot of a
large number of scalar particles and gravitons, as they make their way
out to infinity.  Near the final surface, then, the coupling in
Eqs.(3.20,25) between the linearised perturbations $\phi^{(1)}$ and
${\,}h^{(1)}_{\mu\nu}$ will almost have disappeared.  The perturbed 
scalar field equation at late times is simply
$$\nabla^{\mu}\nabla_{\mu}{\,}\phi^{(1)}{\;}{\,}={\;}{\,}0{\quad},\eqno(4.3)$$
with respect to the spherically-symmetric background geometry
$\gamma_{\mu\nu}{\,}$.

Making the mode decomposition (4.1) of $\phi^{(1)}$, one obtains the
$(\ell{\,},m)$ mode equation
$$\bigl(e^{(b-a)/2}\partial_{r}\bigr)^{2}R_{\ell m} 
+\bigl(\partial_{\tau}\bigr)^{2}R_{\ell m} 
+{1\over 2}\bigl(\partial_{\tau}(a-b)\bigr)
\bigl(\partial_{\tau}R_{\ell m}\bigr)
-V_{\ell}\bigl(\tau{\,},r\bigr)R_{\ell m}{\;}{\,}={\;}{\,}0{\quad}.
\eqno(4.4)$$
\noindent
Here,
$$V_{\ell}(\tau,r){\;}{\,}={\;}{\,}{{e^{b(\tau,r)}\over{r^{2}}}}
{\biggl[\ell(\ell +1)+{{2m(\tau{\,},r)}\over{r}}\biggr]}\eqno(4.5)$$
\noindent
and ${\,}m(\tau{\,},r)$ is defined by 
$$\exp\bigl(-a(\tau{\,},r)\bigr){\;}{\,}
={\;}{\,}1-{{2m(\tau{\,},r)}\over{r}}{\quad}.\eqno(4.6)$$
\noindent
In an exact Schwarzschild solution with no scalar field, one would
have ${\;}e^{b}=e^{-a}=1-{{2M}\over r}{\,}$, with $M$ the Schwarzschild
mass; in that case, $m(\tau,r)$ would be identically $M{\,}$.  The
potential $V_{\ell}(\tau{\,},r)$ of Eq.(4.5) generalises the well-known
massless-scalar effective potential in the exact Schwarzschild geometry
[24], which vanishes at the event horizon $\{r=2M\}$ and at
spatial infinity, and has a peak near $\{r=3M\}$.

The definition (4.6) of $m(\tau{\,},r)$ is also consistent with the usual
description of the Lorentzian-signature Vaidya metric [21,22].  In
terms of a null coordinate $u$ and an intrinsic radial coordinate $r{\,}$,
the Vaidya metric reads
$$ds^{2}{\;}{\,} 
={\;}{\,}-{\,}2{\,}du{\,}dr{\,}-\biggl(1-{{2m(u)}\over r}\biggr)du^2{\,} 
+{\,}r^2{\,}\bigl(d\theta^{2}+\sin^{2}\theta{\;}d\phi^2\bigr){\quad}.
\eqno(4.7)$$
\noindent
Here, $m(u)$ is a monotonic-decreasing, but otherwise freely
specifiable smooth function of $u$, corresponding to a suitable
spherically-symmetric outflow of null particles, for example the
energy-momentum tensor of a black hole evaporating {\it via} emission of
scalar particles at the speed of light.  The Vaidya metric has been
used often to give an approximate gravitational background for
black-hole evaporation at late times [4,5,62].  This connection 
will be treated more thoroughly in our subsequent work.

There is, of course, an analogous decoupled harmonic decomposition,
valid near the final surface, for the weak gravitational-wave
perturbations about the spherically-symmetric background  -- again
described in subsequent work.  For simplicity of exposition, we shall 
in the following Paper II restrict attention to weak-field final
configurations for spin-0 (scalar), and calculate their quantum
amplitudes on the further assumption that the final 3-metric
$h_{ijF}$ is exactly spherically symmetric (in addition to the assumed
spherical symmetry of the initial data $\phi_I$ and $h_{ijI}$).  Once 
the methods are established in the simplest spin-$0$ case, 
generalisation to the case of higher-spin fields becomes relatively 
straightforward.
\end{section}

\begin{section}{The classical action}
Consider, for definiteness, an asymptotically-flat
Lorentzian-signature classical solution $(g_{\mu\nu}{\,},\phi)$ of the
coupled Einstein/massless-scalar field equations, between an initial
hypersurface $\Sigma_I$ and a final hypersurface $\Sigma_F{\,}$, 
separated by a proper Lorentzian time $T$ at spatial infinity.  Write
$S$ for the Lorentzian action functional, which corresponds to the
Riemannian action functional $I$ of Eq.(3.4) with suitable boundary
contributions [15], appropriate to fixing the boundary data
$(h_{ij}{\,},\phi)_I$ and $(h_{ij}{\,},\phi)_F{\,}$, according to 
${\,}iS=-I{\,}$.  Then, at the Lorentzian-signature solution above, 
one has [15,63] the classical action
$$\eqalign{S_{\rm class}\biggl[(h_{ij}{\,},\phi)_{I};(h_{ij}{\,},\phi)_{F};
T\biggr]{\;}{\,}
={\;}{\,}{{1}\over{32\pi}}&\biggl(\int_{\Sigma_{F}} 
-\int_{{\Sigma_I}}\biggr){\,}d^{3}x{\;}{\,}\pi^{ij}{\,}h_{ij}\cr
&+{\,}{{1}\over{2}}{\,}\biggl(\int_{\Sigma_{F}}-\int_{\Sigma_{I}}\biggr){\,}
d^{3}x{\;}{\,}\pi_{\phi}{\;}\phi{\;}-{\,}M{\,}T{\quad}.\cr}\eqno(5.1)$$
\noindent
Here, $\pi^{ij} = \pi^{ji}$ is $(16\pi)$ times the Lorentzian momentum 
conjugate to the 'coordinate' variable $h_{ij}$ on a space-like
hypersurface, in a $3+1$ Hamiltonian decomposition of the
Einstein/massless-scalar theory [64].  Explicitly, in terms of
the Lorentzian-signature second fundamental form $K_{ij} = K_{ji}$ of
the hypersurface [15,49], $\pi^{ij}$ is given by
$$\pi^{ij}{\;}{\,}={\;}{\,}h^{1/2}{\,}(K^{ij}{\,}-K{\,}h^{ij}){\quad},
\eqno(5.2)$$
\noindent
where ${\,}h ={\rm det}(h_{ij})$ and ${\,}K=h^{ij}K_{ij}{\,}.$  
Further, $\pi_{\phi}$ is the Lorentzian momentum conjugate to the 
'coordinate' variable $\phi{\,}$.  Explicitly,
$$\pi_{\phi}{\;}{\,}
={\;}{\,}h^{1\over 2}{\,}n^{\mu}(\nabla_{\mu}\phi){\quad},\eqno(5.3)$$ 
\noindent
where $n^\mu$ denotes the (Lorentzian-signature) future-directed unit
time-like vector normal to the hypersurface.

Suppose instead that one has a complex or a Riemannian solution
$(g_{\mu\nu}{\,},\phi)$ between asymptotically-flat boundary data
$(h_{ij}{\,},\phi)_I$ and $(h_{ij}{\,},\phi)_F$ on initial and final
hypersurfaces $\Sigma_{I}{\,},\Sigma_{F}{\,}$, where the time-separation $T$
at infinity has the form $T=\tau{\,}\exp(-i\theta){\,}$, as in
Eq.(2.3), where $\tau$ is positive real and
${\,}0<\theta\leq\pi/2{\,}$.  This, as above, is expected to provide 
the most natural arena for asymptotically-flat boundary-value problems 
involving gravitation, if strong ellipticity holds, up to gauge.  For such a
solution, the Lorentzian-signature classical action $S_{\rm class}$ 
continues to be defined by Eq.(5.1).  This will in general be complex, 
although for a real Riemannian solution with ${\,}\theta=\pi/2{\,}$, 
the Riemannian action $I_{\rm class}{\,}$, defined by 
${\,}I_{\rm class}=-{\,}i{\,}S_{\rm class}{\,}$, will be real.
The boundary contribution at spatial infinity to the Riemannian action
functional $I$ corresponding to Eq.(5.1) is ${\,}+M\tau$ [15].  
The boundary contributions to the functional $I{\,}$, due to the 
presence of the boundaries $\Sigma_I$ and $\Sigma_F$ with data 
$(h_{ij}{\,},\phi)_I$ and $(h_{ij}{\,},\phi)_F$ specified on them, are 
$$I_I{\,}+{\,}I_F{\;}{\,} 
={\;}{\,}{{1}\over{32\pi}}{\;}\biggl(\int_{\Sigma_I}-\int_{\Sigma_F}\biggr)
{\,}d^{3}x{\,}{~}_{e}\pi^{ij}{\;}h_{ij}{\,}
+{\,}{1\over 2}{\,}\biggl(\int_{\Sigma_F}-\int_{\Sigma_I}\biggr){\,} 
d^{3}x{\,}{~}_{e}\pi_{\phi}{\;}\phi{\quad}.\eqno(5.4)$$
\noindent
Here,
$${~}_e\pi^{ij}{\;}{\,} 
={\;}{\,}h^{1/2}{\;}\bigl({~}_{e}K^{ij}-{~}_{e}K{\;}h^{ij}\bigr)\eqno(5.5)$$
\noindent
is given by the same formula as $\pi^{ij}$ in Eq.(5.2), except that
$K_{ij}$ has been replaced by the 'Euclidean' second fundamental form
${~}_{e} K_{ij}{\,}$, as defined and used in Eqs.(2.6.23,24) of [15].  
In particular, 
$$_{e}K_{ij}{\;}{\,}={\;}{\,}-{\,}i{\,}K_{ij}{\quad}.\eqno(5.6)$$
Similarly, the scalar-momentum variable $\pi_{\phi}$ of Eq.(5.3) has
been replaced by its 'Euclidean' version $_{e}\pi_\phi{\,}$, defined by 
$${~}_{e}\pi_{\phi}{\;}{\,} 
={\;}{\,}h^{1/2}{~}_{e}n^{\mu}{\;}(\nabla_{\mu}\phi){\quad},\eqno(5.7)$$
\noindent
where [15]
$${~}_{e}n^{\mu}{\;}{\,}={\;}{\,}-{\,}i{\,}n^{\mu}\eqno(5.8)$$
\noindent
denotes the unit future-directed Riemannian normal.\par
\smallskip
\indent
The quantity $M$ in Eq.(5.1) denotes the Arnowitt-Deser-Misner (ADM)
mass of the 'space-time', as measured near spatial infinity from the
$({1/r})$ part of the fall-off of the intrinsic spatial metric
$h_{ij}$ on $\Sigma_I$ and $\Sigma_F$ [23,24].  As mentioned in
Section 5, it is essential, for a well-posed asymptotically-flat
boundary-value problem, that the intrinsic metrics $h_{ijI}$ and
$h_{ijF}$ be chosen to have the same value of $M$.  Otherwise, if
$\,M_{I} \neq M_F\,$, then any classical infilling 'space-time' will have
$\Sigma_I$ and $\Sigma_F$ badly embedded near spatial infinity, and
the entire 4-metric $g_{\mu\nu}$ will not fall off as rapidly as it
should, as ${\,}r\rightarrow \infty$ [61].

In applications to black-hole particle emission, we naturally make use 
of the perturbative splitting 
${\;}g_{\mu\nu}=\gamma_{\mu\nu}+h^{(1)}_{\mu\nu}+\ldots{\;},
{\quad}\phi=\Phi+\phi^{(1)}+\ldots{\;},{\;}{\,}$
where the spherically-symmetric 'background' 
$(\gamma_{\mu\nu}{\,},\Phi)$ obeys the coupled
Einstein/massless-scalar classical field equations, as does the full 
classical solution $(g_{\mu\nu}{\,},\phi){\,}$.  (The formal device of 
including a small parameter $\epsilon$ has been relaxed here:  
we now set ${\,}\epsilon{\,}={\,}1{\,}$.)

The linearised fields $h^{(1)}_{\mu\nu}$ and $\phi^{(1)}$ may be 
decomposed into sums of appropriate angular harmonics labelled by 
quantum numbers $(\ell{\,},m)$ as in Section 5, and without loss of 
generality it may be assumed that any spherically-symmetric ${\,}{\ell}=0$ 
linear-order perturbation modes have been absorbed into the 
spherically-symmetric background $(\gamma_{\mu\nu}{\,},\Phi){\,}$.  
Then (say) the Lorentzian classical action $S_{\rm class}$ of Eq.(5.1) 
may be split as
$$S_{\rm class}{\;}{\,} 
={\;}{\,}S^{(0)}_{\rm class}+S^{(2)}_{\rm class}+S^{(3)}_{\rm class}{\;} 
+{\;}\ldots{\quad}.\eqno(5.9)$$
\noindent
Here, $S^{(0)}_{\rm class}$ is the background action, given by
Eq.(5.1), but evaluated for the spherically-symmetric solution
$(\gamma_{\mu\nu}{\,},\Phi){\,}.$  The mass $M$ appearing in 
$S^{(0)}_{\rm class}$ will be that determined from $(\gamma_{ij})_{I}$ or 
$(\gamma_{ij})_{F}{\,}$.  The next term is 
${\,}S^{(2)}_{\rm class}{\,}$, formed quadratically from the
linear-order perturbations; one may verify that the linear-order term
$S^{(1)}_{\rm class}$ is zero, because of the above definitions.  In
an obvious notation, one has 
$$S^{(2)}_{\rm class}{\;}{\,} 
={\;}{\,}{{1}\over{32\pi}}{\,}
\biggl(\int_{\Sigma_F}-\int_{\Sigma_I}\biggr){\,}d^{3}x{\;}{\,}
\pi^{(1)ij}{\;}h^{(1)}_{ij}{\,}
+{\,}{1\over 2}{\,}\biggl(\int_{\Sigma_F}-\int_{\Sigma_I}\biggr)
{\,}d^{3}x{\;}{\,}\pi^{(1)}_{\phi}{\;}\phi^{(1)}{\quad}.\eqno(5.10)$$ 
\noindent
Note that there is no contribution to the second-order expression
$S^{(2)}_{\rm class}$ from the $-MT$ term in Eq.(5.1), again because
of the above definitions.

The expression (5.1) for 
$S_{\rm class}[(h_{ij}{\,},\phi)_{I}{\,};(h_{ij}{\,},\phi)_{F}{\,};T]$, 
together with the asymptotic series (5.9) for the classical action and the
expression (5.10) for $S^{(2)}_{\rm class}$ formed from the linearised 
perturbations, will be basic in calculations concerning quantum 
amplitudes in subsequent work.
\end{section}

\begin{section}{Conclusion}
This paper has been concerned with setting up a basic framework and
formalism for treating quantum amplitudes involving possibly strong
gravitational fields, governed by a Lagrangian containing a locally
supersymmetric version of Einstein gravity and matter.  This includes 
the case of gravitational collapse to a black hole, but is
considerably more general, being applicable also to many cosmological 
problems involving small fluctuations of an isotropic homogeneous 
universe, in the quantum context.

The underlying approach in this paper has been to calculate the
quantum amplitude to go from data (both for gravity and any matter
fields) specified on an initial spacelike hypersurface $\Sigma_I$ to
corresponding data given on a final hypersurface $\Sigma_F{\,}$.  Since,
in the black-hole context mainly studied here, the space-time should
be asymptotically flat, we take the simplest case in which both
$\Sigma_I$ and $\Sigma_F$ are diffeomorphic to Euclidean space 
${\Bbb R}^{3}$, and such that their intrinsic 3-dimensional metrics
$h_{ijI}{\,},{\,}h_{ijF}$ are asymptotically flat at spatial infinity.  
The 'boundary data',  which should determine the quantum amplitude 
uniquely, then consist of $h_{ij}$ and suitable components of any
matter fields, on $\Sigma_I$ and $\Sigma_F{\,}$, together with the 
(Lorentzian) proper time-interval $T$ between $\Sigma_I$ and 
$\Sigma_F{\,}$, measured near spatial infinity.  Following Feynman's 
$+i\epsilon$ prescription [13], we rotate $T$ into the complex: 
$T\rightarrow{\mid}T{\mid}\exp(-i\theta){\,}$, 
for ${\,}0<\theta\leq\pi/2{\,}$.  One expects that the corresponding 
complex {\it classical} boundary-value problem for the 
Einstein/matter field equations should be well-posed, unlike the 
purely Lorentzian case with $T$ real.  The remaining analysis is
mainly concerned with properties and consequences of such complex 
solutions of the field equations, where the data $h_{ij}{\,},$ etc., 
for gravity and matter are held fixed on the boundaries $\Sigma_I$ and
$\Sigma_F{\,}$, but with 
$T\rightarrow{\mid}T{\mid}\exp(-{\,}i\theta){\,}$.
The quantum amplitude for linearised perturbations is given principally
through the second-variation classical action $S^{(2)}_{\rm class}{\,}$,
as a functional of the boundary data, with the amplitude proportional
to ${\,}\exp(i{\,}S^{(2)}_{\rm class}){\,}$, except near Planckian energies.
Feynman's prescription then requires that we take the limit 
${\,}\theta{\,}{\rightarrow}{\,}0_{+}$ to obtain the Lorentzian 
amplitude.

In the following Paper II of this series, we shall evaluate the above
amplitude for a model with Einstein gravity and minimally-coupled
scalar field $\phi{\,}$, in the case that the perturbations on $\Sigma_F$
are only in $\phi{\,}$, but not in the final gravitational data 
$h_{ij}{\,}$, which are there taken to be spherically symmetric.  
In particular, this is applicable to the case of spin-0 (scalar) 
radiation from gravitational collapse to a black hole.  In further
work, this description will be related to the alternative language 
of Bogoliubov transformations, in which much of the earlier work on black-hole 
evaporation was cast [3,34,35].  Next, we will analyse in some detail 
the semi-classical description of the region of the space-time 
containing the outgoing flux of (here, spin-0 and spin-2) radiation, 
with the help of the Vaidya metric [21].  Yet another kind of 
'transformation' can also be analysed, namely, that between the
language of this work and the language of coherent and squeezed
states; this approach makes it easier to take an overview and see 
(for example) the similarity between our 'local-collapse' or 
'black-hole' description and a cosmological version, in which the only 
boundary is (say) a compact 3-surface, such as a 3-sphere 
$S^{3}$ [15,36].  The calculations of Paper II for the quantum 
amplitude for purely scalar-field (spin-0) emission will be 
generalised to the cases of spin-1 Maxwell (or Yang-Mills) radiation 
and to spin-2 graviton (gravitational-wave) emission.  The fermionic 
case of a massless $s={1\over 2}$ (neutrino) field will also be treated.  
Work on further aspects of the approach, as outlined in the
Introduction and in [11,33], has also been carried out and should 
soon be completed.
\end{section}

\begin{section}*{Acknowledgments}
We are grateful to Jim Hartle for invaluable discussions
about this work. We also owe warm thanks to Ian Drummond for 
encouragement and advice at two crucial stages in the genesis of the 
thesis. Finally, very warm gratitude to K.M. Wheeler for much 
useful help in the final stages.
\end{section}

\begin{section}*{References}

\everypar{\hangindent\parindent}

\noindent [1] M.K.Parikh and F. Wilczek, 
Phys. Rev. Lett. {\bf 85}, 5042 (2000).

\noindent [2] M.Parikh, 
Gen. Relativ. Gravit., {\bf 36}, 2419 (2004).

\noindent [3] S.W.Hawking, 
Nature (London) {\bf 248}, 30 (1974);
Phys. Rev. D {\bf 14}, 2460 (1976);
Commun. Math. Phys. {\bf 43}, 199 (1975).

\noindent [4] P.H{\'a}j{\'i}{\v c}ek and W.Israel, 
Phys. Lett. A {\bf 80}, 9 (1980).

\noindent [5] J.Bardeen, 
Phys. Rev. Lett. {\bf 46}, 382 (1981).

\noindent [6] S.W.Hawking, 
Commun. Math. Phys. {\bf 87}, 395 (1982).

\noindent [7] S.W.Hawking, `Boundary Conditions of the Universe' 
in {\it Astrophysical Cosmology}, 
Proceedings of the Study Week on Cosmology and Fundamental Physics, 
eds. H.A.Br{\"u}ck, G.V.Coyne and M.S.Longair. 
Pontificia Academiae Scientarium: Vatican City, {\bf 48}, 563 (1982).

\noindent [8] R.M.Wald, in {\it Quantum Theory of Gravity}, 
ed. S.Christensen, (Adam Hilger, Bristol) 160 (1984).

\noindent [9] A.N.St.J.Farley, 
'Quantum Amplitudes in Black-Hole Evaporation', 
Cambridge Ph.D. dissertation, approved 2002 (unpublished).

\noindent [10] S.W.Hawking, 
communication, GR17 Conference, Dublin, 18-24 July (2004).

\noindent [11] A.N.St.J.Farley and P.D.D'Eath, 
Phys Lett. B {\bf 601}, 184 (2004).

\noindent [12] E.P.S.Shellard, 
'The future of cosmology:  observational and computational prospects', 
in {\it The Future of Theoretical Physics and Cosmology}, 
eds. G.W.Gibbons, E.P.S.Shellard and S.J.Rankin
(Cambridge University Press, Cambridge) 755 (2003).

\noindent [13] R.P.Feynman and A.R.Hibbs, 
{\it Quantum Mechanics and Path Integrals}, 
(McGraw-Hill, New York) (1965).

\noindent [14] P.R.Garabedian, 
{\it Partial Differential Equations}, (Wiley, New York) (1964).

\noindent [15] P.D.D'Eath, 
{~}{\it Supersymmetric {~}Quantum {~}Cosmology}, 
{~}(Cambridge {~}University {~}Press, {~}Cambridge) (1996).

\noindent [16] W.McLean, 
{\it Strongly Elliptic Systems and Boundary Integral Equations}, 
(Cambridge University Press, Cambridge) (2000); 
O.Reula, 'A Configuration space for quantum gravity and
solutions to the Euclidean Einstein equations in a slab region',
Max-Planck-Institut f\"ur Astrophysik, {\bf MPA}, 275 (1987).

\noindent [17] D.Christodoulou, Commun. Math. Phys. {\bf 105} 337 (1986); 
{\bf 106} 587 (1986); {\bf 109} 591, 613 (1987);
Commun. Pure Appl. Math. {\bf 44}, 339 (1991); {\bf 46}, 1131 (1993).

\noindent [18] P.D.D'Eath and A.Sornborger, 
Class. Quantum Grav. {\bf 15}, 3435 (1998).

\noindent [19] P.D.D'Eath, 
'Numerical and analytic estimates for Einstein/scalar boundary-value
problems', in progress.

\noindent [20] A.Das, M.Fischler and M. Ro\v cek, 
Phys.Lett. B {\bf 69}, 186 (1977).

\noindent [21] P.C.Vaidya,  
Proc. Indian Acad. Sci. {\bf A 33}, 264 (1951).

\noindent [22] R.W.Lindquist, R.A.Schwartz and C.W.Misner, 
Phys Rev.{\bf 137}, 1364 (1965).

\noindent [23] R.Arnowitt, S.Deser and C.W.Misner, 
'Dynamics of General Relativity', 
in {\it Gravitation: An Introduction to Current Research}, 
ed. L.Witten (Wiley, New York) (1962).

\noindent [24] C.W.Misner, K.S.Thorne and J.A.Wheeler, 
{\it Gravitation}, (Freeman, San Francisco) (1973).

\noindent [25] J.Wess and J.Bagger, 
{\it Supersymmetry and Supergravity}, 2nd. edition, 
(Princeton University Press, Princeton) (1992).

\noindent [26] P.D.D'Eath, 
'Loop amplitudes in supergravity by canonical quantization', in 
{\it Fundamental Problems in Classical, Quantum and String Gravity}, 
ed. N.S{\'a}nchez (Observatoire de Paris) 166 (1999); hep-th/9807028.

\noindent [27] P.D.D'Eath, 
'What local supersymmetry can do for quantum cosmology', 
in {\it The Future of Theoretical Physics and Cosmology}, 
eds. G.W.Gibbons, E.P.S.Shellard and S.J.Rankin 
(Cambridge University Press, Cambridge) 693 (2003).

\noindent [28] S.W.Hawking, Phys.Lett.B {\bf 126}, 175 (1983).

\noindent [29] G.V.M.Esposito, 
{\it Quantum Gravity, Quantum Cosmology and Lorentzian Geometries}, 
Lecture Notes in Physics m12 (Springer, Berlin) (1994).

\noindent [30] A.N.St.J.Farley and P.D.D'Eath, 
Class. Quantum Grav. {\bf 22}, 3001 (2005).

\noindent [31] L.D.Faddeev and A.A.Slavnov, {\it Gauge Fields},
(Benjamin-Cummings, Reading, Mass.) (1980).

\noindent [32] D.Bao, Y.Choquet-Bruhat, J.Isenberg and P.B.Yasskin, 
J. Math. Phys. {\bf 26}, 329 (1985).

\noindent [33] A.N.St.J.Farley and P.D.D'Eath, 
Phys. Lett B {\bf 613}, 181 (2005).

\noindent [34] N.D.Birrell and P.C.W.Davies, 
{\it Quantum fields in curved space}, 
(Cambridge University Press, Cambridge) (1982).

\noindent [35] V.P.Frolov and I.D.Novikov, {\it Black Hole Physics}, 
(Kluwer Academic, Dordrecht) (1998).

\noindent [36] J.B.Hartle and S.W. Hawking, 
Phys. Rev. D {\bf 28} 2960 (1983).

\noindent [37] P.A.M.Dirac, {\it Lectures on Quantum Mechanics}, 
(Academic Press, New York) (1965).

\noindent [38] P.D.D'Eath, {\it Black Holes: Gravitational Interactions}, 
(Oxford University Press, Oxford) (1996).

\noindent [39] G. 't Hooft, Phys. Lett. B {\bf 198}, 61 (1987).

\noindent [40] S. Giddings, 'Black holes at accelerators',
in {\it The Future of Theoretical Physics and Cosmology}, 
eds. G.W.Gibbons, E.P.S.Shellard and S.J.Rankin 
(Cambridge University Press, Cambridge) 278 (2003).

\noindent [41] A.N.St.J.Farley and P.D.D'Eath, 
Class. Quantum Grav. {\bf 22}, 2765 (2005).

\noindent [42] A.N.St.J.Farley and P.D.D'Eath, 
'Spin-3/2 Amplitudes in Black-Hole Evaporation', 
in progress.

\noindent [43] J.Mathews, J. Soc. Ind. Appl. Math. {\bf 10}, 768 (1962).

\noindent [44] J.N.Goldberg, A.J.MacFarlane, E.T.Newman, F.Rohrlich and 
E.C.G.Sudarshan, J. Math. Phys. {\bf 8}, 2155 (1967).

\noindent [45] S.Kobayashi and K.Nomizu, 
{\it Foundations of Differential Geometry}, 
Vol.II (Wiley, New York) (1969).

\noindent [46] R.Geroch, Commun. Math. Phys. {\bf 13}, 180 (1969).

\noindent [47] M.W.Choptuik,
'''Critical'' Behaviour in Massless Scalar Field Collapse', 
in {\it Approaches to Numerical Relativity},
ed. R.d'Inverno (Cambridge University Press, Cambridge) (1992).

\noindent [48] M.W.Choptuik, Phys. Rev. Lett. {\bf 70}, 9 (1993).

\noindent [49] S.W.Hawking and G.F.R.Ellis, 
{\it The large scale structure of space-time}, 
(Cambridge University Press, Cambridge) (1973).

\noindent [50] U.H.Gerlach and U.K.Sengupta, 
Phys. Rev. D {\bf 18}, 1789 (1978).

\noindent [51] D.Brill and J.B.Hartle,  
Phys. Rev. {\bf 135}, 1327 (1964).

\noindent [52] R.Isaacson, Phys. Rev. {\bf 166}, 1263, 1272 (1968).

\noindent [53] A.Nayfeh, {\it Perturbation Methods}, 
(Wiley-Interscience, New York) (1973).

\noindent [54] C.M.Bender and S.A.Orszag, 
{\it Advanced Mathematical Methods for Scientists and Engineers} 
(Springer, New York) (1999).

\noindent [55] R.d'Inverno, {\it Introducing Einstein's Relativity}
(Oxford University Press, Oxford) (1992).

\noindent [56] J.D.Jackson, {\it Classical Electrodynamics},
(Wiley, New York) (1975).

\noindent [57] T.Regge and J.A.Wheeler, 
Phys. Rev. {\bf 108}, 1063 (1957).

\noindent [58] C.V.Vishveshwara, 
Phys. Rev. D {\bf 1}, 2870 (1970).

\noindent [59] F.J.Zerilli, 
Phys. Rev. D {\bf 2}, 2141 (1970).

\noindent [60] J.A.H.Futterman, F.A.Handler and R.A.Matzner,   
{\it Scattering from Black Holes} 
(Cambridge University Press, Cambridge) (1988).

\noindent [61] J.A.Wheeler, 
'Superspace and the Nature of Quantum Geometrodynamics' 
p.303, in {\it Battelle Rencontres}, 
ed. C.M.DeWitt and J.A.Wheeler (W.A.Benjamin, New York) (1968).

\noindent [62] W.A.Hiscock, 
Phys. Rev D {\bf 23}, 2813, 2823 (1981).

\noindent [63] G.W.Gibbons and S.W.Hawking, 
Phys. Rev. D {\bf 15}, 2738 (1977).

\noindent [64] J.J.Halliwell and S.W.Hawking, 
Phys. Rev. D {\bf 31}, 1777 (1985).

\end{section}

\end{document}